# Helical Channel Design and Technology for Cooling of Muon Beams

K. Yonehara[a], Y. S. Derbenev[b] and R. P. Johnson[c]

[a]Fermi National Accelerator Laboratory
P.O. Box 500, Batavia, IL 60510, USA
[b]Thomas Jefferson National Accelerator Facility
12000 Jefferson Ave., Newport News, VA 23606, USA
[c]Muons, Inc.
552 N. Batavia Ave., Batavia, IL 60510, USA

**Abstract.** Novel magnetic helical channel designs for capture and cooling of bright muon beams are being developed using numerical simulations based on new inventions such as helical solenoid (HS) magnets and hydrogen-pressurized RF (HPRF) cavities. We are close to the factor of a million six-dimensional phase space (6D) reduction needed for muon colliders. Recent experimental and simulation results are presented.

## INTRODUCTION

A muon collider is suggested for a future lepton collider machine in the P5 report in 2008 [1]. Because a muon has 210 times the mass of the electron, its synchrotron radiation is enormously reduced and it can be accelerated and stored in a circular ring. Therefore, the size of the accelerator complex for muon collider can be much smaller than for a CLIC type machine with the same center of mass energy (com). The Muon Accelerator Program (MAP) was established in 2009 [2] to develop muon beam cooling and acceleration technology. Table 1 shows the parameter list of present and future collider machines [3]. The primary goal in the MAP is 1.5 TeV com and $10^{34}$ cm$^{-2}$s$^{-1}$ which are comparable with CLIC parameters. Unlike the CLIC machine, a muon collider has the capability of an energy upgrade without changing the size of facility by increasing magnetic fields and increasing the number of passes through a recycling linear accelerator (RLA), similar to CEBAF at JLab.

**TABLE 1.** Parameters of present and future collider machines

|  | Species | $E_{com}$ | Size | number of elements | Cost | Operation time |
|---|---|---|---|---|---|---|
| Unit |  | TeV | Km |  |  |  |
| LHC | p-p | 14 | D 8.4 | 11,000 | US $4.6 B | 2009-present |
| ILC | e+-e- | 0.5 | L 30 | 38,000 | US $8.0 B in 2007 | Unknown |
| CLIC | e+-e- | 1.5-3.0 | L 50 | 260,000 | Estimate in 2010 | Unknown |
| Muon Collider | μ+-μ- | 1.5-4.0 | D 2.0 | 10,000 | Unknown | Unknown |

Muons are generated from pion decays. So the initial 6D emittance of a muon beam is too large to be accelerated by any conventional RF acceleration system. Since muon 6D cooling must be achieved within the muon's short lifetime (2.2γ μsec, where γ is the usual relativistic parameter) the only technique applicable for muon beam phase space cooling is ionization cooling. To maximize the cooling efficiency, a helical cooling channel (HCC) filled with a homogeneous ionization cooling absorber is proposed [4]. The HCC consists of helical dipole, helical quadrupole, and solenoidal magnetic fields components configured to produce continuous dispersion. Therefore, a higher (lower) momentum muon traverses longer (shorter) path length in the helical magnet. By using a hydrogen gas as a homogeneous ionization cooling absorber, emittance exchange takes place and the 6D emittance of the muon beam is cooled down continuously along with the channel length. RF cavities are superimposed in the HCC to regain the ionization energy lost in cooling absorbers. Recently, the first end-to-end 6D cooling simulation has been made in a 300-m-long HCC. The 6D phase space is successively reduced by a factor $10^5$ in the channel which is close to minimum requirement for muon collider design ($10^6$).

## SIMULATIONS OF HCC DESIGNS WITH HOMOGENEOUS ENERGY ABSORBER

Figure 1 shows the first end-to-end homogeneous gaseous hydrogen filled HCC cooling simulation. As the beam is cooled, smaller HCC dimensions are possible so that the magnetic field strength and the RF cavity frequency can each be increased along the channel length. The stronger magnetic field improves the ionization cooling and higher frequency RF improves efficiency. The helical field parameters are listed in Table 2. The average initial and final transverse emittances are 20.4 and 0.34 mm rad, while the initial and final longitudinal emittances are 42.8 and 1.1 mm, respectively. Hence the transverse and longitudinal cooling factors are 60 and 40, and the obtained 6D cooling factor is 60×60×40=144,000 > $10^5$. This is close to the minimum requirement for muon collider designs ($10^6$) [5].

As shown in Figure 1, extra cooling channels can be used to obtain additional cooling factors. Three examples are represented in Figure 1. The High Temperature Superconducting (HTS) Solenoid channel uses a very high solenoid field, e.g. 40 to 50 Tesla, to reach an extremely low transverse emittance [6]. Another technique is called Parametric resonance Ionization Cooling (PIC), in which beam in the PIC channel is excited by a half integer resonance. By putting ionization cooling material at the largest angular divergence position, extra transverse cooling takes place. The Reverse EMittance EXchange (REMEX) technique uses wedge absorbers to reduce the transverse emittance at the expense of the longitudinal emittance up to the point that the bunch length gets long enough to impact the collider luminosity due to the hour glass effect. PIC and REMEX can be made in a double helical structure, discussed in Ref. [7].

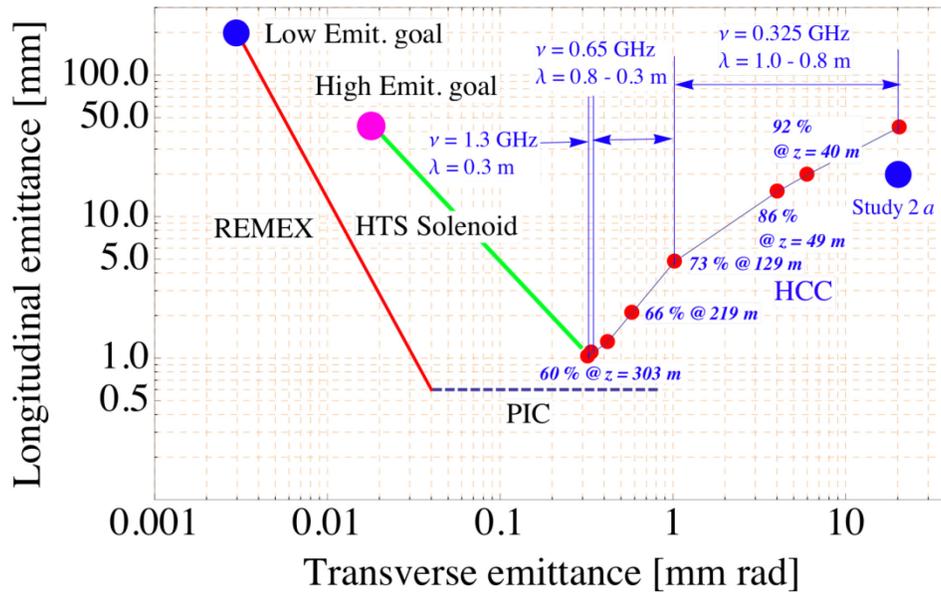

**FIGURE 1.** Emittance evolution in homogeneous absorber filled helical cooling channel. The simulation has been done using analytical field expressions with parameters shown in Table 2. The muon beam phase space point called Study2a is the end of a conventional frontend channel. The HCC acceptance is much larger than the Study 2a emittance from neutrino factory studies.

The beam lattice in the simulations uses analytical field expressions based on Bessel functions. The lattice parameters are listed in Table 2. Taylor expansions up to second order are used in the analytical field. Consequently, the analytical helical magnetic field has a big non-linearity near the boundary of stable beam phase space. To compensate the beam instability due to the non-linear lattice component, a very high RF gradient, 27 MV/m was used in the simulation. It requires a peak RF power of 10 MW/m at ν = 325 MHz and 60 Hz repetition rate for a 1.5 TeV com collider. From recent helical design studies, the beam instability due to non-linear fields is greatly reduced by using the HS magnet, described below. In this case, the RF gradient can be reduced to 17 MV/m. If the RF cavities operate at cryogenic temperature, the resistivity can be significantly lowered to reduce the peak power requirements. By taking into account all these improvements, the required peak RF power of about 0.2 MW/m seems to be feasible but still high. Further investigations will be needed to reduce the required RF power or to find new power sources.

**TABLE 2.** Lattice parameters in a series of HCCs. The lattice is designed for μ+ beam with average momentum 200 MeV/c. The tangent of the pitch angle is 1.0.

| | Length | Helical dipole | Helical quadrupole | Solenoid | Helical period | RF frequency |
|---|---|---|---|---|---|---|
| Unit | m | T | T/m | T | M | MHz |
| 1 | 40 | 1.3 | -0.5 | -4.2 | 1.0 | 325 |
| | 49 | 1.4 | -0.6 | -4.8 | 0.9 | 325 |
| | 129 | 1.7 | -0.8 | -5.2 | 0.8 | 325 |
| 2 | 219 | 2.6 | -2.0 | -8.5 | 0.5 | 650 |
| | 243 | 3.2 | -3.1 | -9.8 | 0.4 | 650 |
| | 273 | 4.3 | -5.6 | -14.1 | 0.3 | 650 |
| 3 | 303 | 4.3 | -5.6 | -14.1 | 0.3 | 1300 |

## HELICAL BEAM ELEMENT STUDIES

The magnets and RF cavities of the HCC have been experimentally investigated. To compensate energy loss in the ionization cooling process, high pressure RF (HPRF) cavities are continuously integrated into the channel. Potential advantages of the continuous HPRF cavities include 1) 6D phase space cooling is continuously generated, 2) the energy absorber and RF cavities occupy the same real estate, and 3) the high dielectric strength of the pressurized hydrogen permits operation with high electric field gradients in strong magnetic fields.

In terms of the field expansions, the HCC appears to have a complicated magnetic field structure. However, the required components are elegantly provided by a simple magnetic structure called a Helical Solenoid made of a series of offset circular coils as shown in Figure 2. The HS has the virtue that it provides a continuous, smoothly-varying field which will not drive resonances since there are not fringe fields as found in magnetic channels with lumped elements. Such a simple magnet structure provides us more freedom. For instance, the compactness of the HCC is sometimes a problem since there is not much space to incorporate HPRF cavities. By using a large bore HS magnet with an external solenoid as shown in the right side of Figure 2, the magnet can have more space to mount HPRF cavities and their infrastructure.

### Helical Solenoid Magnet

The HS magnet has been studied to generate the optimum helical field structure [8]. Figure 2 shows schematic pictures of HS magnets. Pancake coils are located are centered on the helical beam path. Helical dipole and helical quadrupole components are generated by superimposed fields from neighbor coils. Therefore, the strengths of helical dipole and quadrupole are strongly dependent on the HS coil geometry and coil locations. An additional pure solenoid coil is needed to adjust the solenoid component on the beam path as shown in the right side of Figure 2.

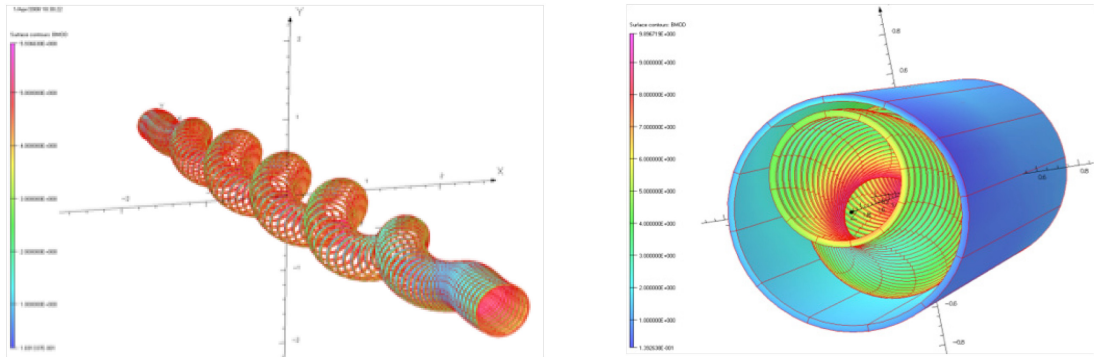

**Figure 2.** Schematic picture of HS magnet. (Left) Bare HS magnet with matching transition sections before and after to a simple solenoid . (Right) HS magnet with a large solenoid correction coil to adjust the solenoid component along the beam path.

Two HCC sections are chosen to demonstrate the feasibility of the HS magnet. Table 3 shows the geometry and current density of sampled HS magnets. Note that the inner radius of HS coil is larger than the radius of the RF cavity to incorporate it and its infrastructure into the HS magnet. It is also worth to mention that there is a gap between HS coils to put the RF power cable through the gap. The current density must be lower than the critical current density in a $NiSn_3$ wire with a 20 % safety margin. There are two remarkable observations in the HS coil design. First, the helical field with analytical field expressions can be reproduced by the HS magnet even if the HS coil bore is extremely large to integrate RF cavity. Second, since the helical field in the HS magnet is much more uniform than the analytical field, the acceptable momentum range increases about 20 %. It means that beam instabilities due to non-linear fields is significantly suppressed and the required RF gradient is reduced. End-to-end 6D cooling simulations with fields generated from HS magnet coils rather than Taylor expansions is underway.

**Table 3.** HS coil geometry and current density comparisons to the field parameters with analytical fields based on Bessel functions. $a$ is a radius of the reference orbit, $\lambda$ is the helical period, $\nu$ is the RF frequency, $R_{coil}$ is the coil center position, $R_{in}$ and $R_{out}$ are the inner and outer radius of the HS coil, $l_{coil}$ is the coil length, $N_{coil}$ is the number of coils per helical period, b, b' and bz are helical dipole and quadrupole and solenoid field strengths on the reference orbit, bsol is a correction solenoid field strength on the reference orbit, and I is a current density in the HS coil.

|  | $a$ | $\lambda$ | $\nu$ | $R_{coil}$ | $R_{in}$ | $R_{out}$ | $l_{coil}$ | $N_{coil}$ | b | b' | bz | bsol | I |
|---|---|---|---|---|---|---|---|---|---|---|---|---|---|
| Unit | cm | M | MHz | cm | Cm | cm | Cm |  | T | T/m | T | T | A/m$^2$ |
| 1st HCC | 16 | 1.0 | 325 | 28 | 35 | 40 | 2.5 | 20 | 1.31 | 0.53 | -4.31 | 0.55 | -194 |
| Bessel HCC |  |  |  |  |  |  |  |  | 1.31 | 0.55 | -4.32 |  |  |
| Final HCC | 6.3 | 0.4 | 650 | 16 | 18 | 28 | 1.25 | 16 | 3.26 | 4.56 | -10.7 | 6.73 | -333 |
| Bessel HCC |  |  |  |  |  |  |  |  | 3.22 | 3.14 | -10.6 |  |  |

## High Pressure RF Cavity

A high pressure hydrogen gas filled RF cavity has been demonstrated without beam. The result shows that dense gas allows for high electric field gradients in RF cavities located in strong magnetic fields (Figure 3).

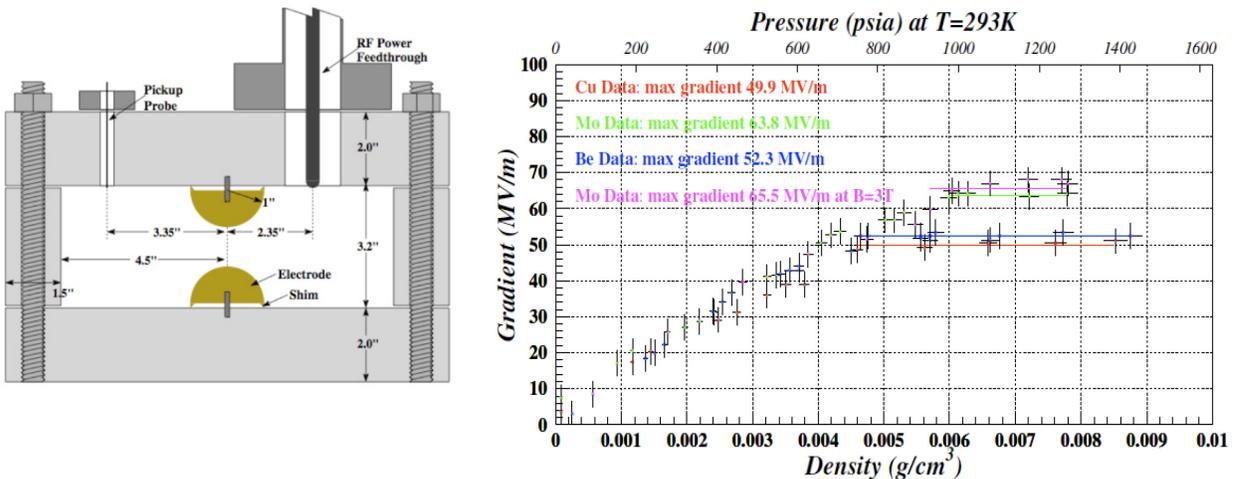

**Figure 3**: (Left) Schematic drawing of cross-sectional view of the HPRF test cell. There are two semispherical electrodes inside the cavity to enhance the electric field in a narrow area. (Right) Maximum electric field as a function of gas pressure. The magenta points correspond to the observed maximum electric field in 3 Tesla solenoid field to be compared to the green points taken with no external magnetic field. No field degradation has been observed in the HPRF cavity.

In addition, an intense muon beam will form an electron swarm in the HPRF cavity via the ionization process. A large amount of RF power may be consumed by the swarm if the recombination rate of the electrons is significantly longer than the RF period. That is, the electrons are mobile enough that they could heat the hydrogen molecules, effectively reducing the Q of the cavity. We have studied the RF breakdown in pressurized cavities without beam to understand electron swarm dynamics from two different approaches [9].

The RF cavity electrically forms a high-Q LC resonance circuit. Therefore, the RF resonant field is very sensitive to an electron swarm which can be generated by a breakdown event and studied using an RF pickup probe. An electrical model of the cavity has been used to extract information about the discharge. Initially, field emission starts a streamer propagating across the cavity which changes into an arc upon reaching across the cavity. Initially the current grows exponentially, but when it reaches a current of several hundred amps a pinch occurs and the current can increase to more than 1000 amps near the end of the discharge. The period before the pinch has been analyzed by the reduction of the cavity voltage versus time. Using the known mobility of the electrons in the gas allows an estimation to be made of the total number of electrons participating in the initial stages of the discharge. A plot of the growth of the number of electrons per half cycle is shown in the left plot of Figure 4. It seems that the RF breakdown starts growing when the total number of electrons reaches $10^{14}$ electrons.

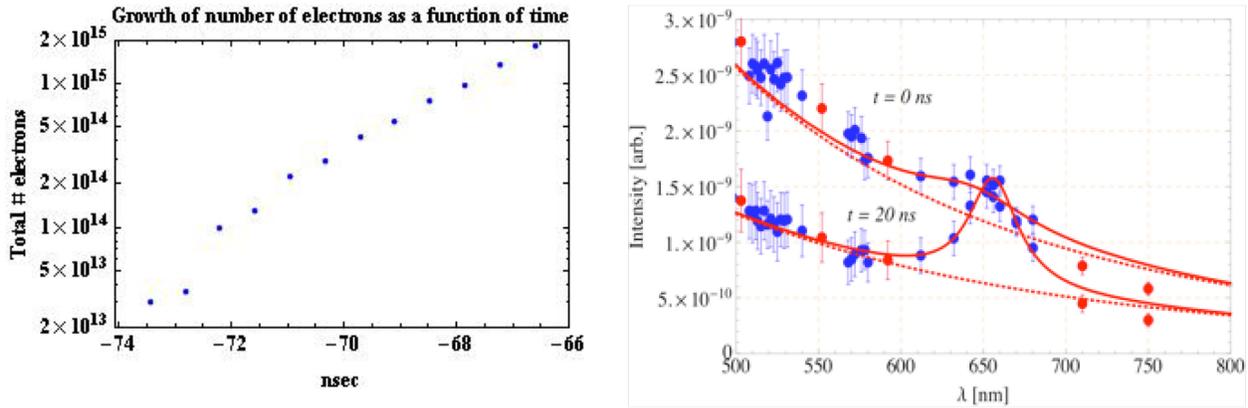

**Figure 4:** (Left) Growth of electron swarm as a function of a half of RF cycles. (Right) Spectra from 740 psi HPRF cavity with various timing. Red points are chosen for the blackbody radiation analysis.

A spectroscopic measurement has been done as another approach to study electron swarm dynamics. Strong light was observed during the RF breakdown. In order to obtain a clear light signal from hydrogen gas, the light was accumulated over 100 RF breakdowns. The applied electric field strength was carefully adjusted to make the breakdown probability of less than 1 % to avoid sequential breakdowns. To obtain a stable trigger signal, an additional PMT and an optical fiber were connected to another optical port. The trigger level was high enough to remove any background signal, e.g. from cosmic rays. The right hand side of Figure 4 shows the spectroscopy of the RF breakdown in the hydrogen gas filled RF cavity. The "zero" timing is at the peak of the spectroscopic light signal. Since only one PMT was assembled on the spectrometer only one wavelength of light was taken for each data acquisition. It is a well-known phenomenon that the resonant light is broadened due to the local field from the electron swarm (Stark effect). The electron density in the swarm can be estimated from the width of the resonance peak. It is typically $10^{18}$ cm$^{-3}$ although the value is varied at different time and different gas pressure. Since the size of the plasma sheath is typically 100 μm, $10^{14}$ electrons contributed to the resonance line broadening. This naïve estimation is consistent with above current flow analysis.

## Beam test in HPRF cavity

The beam test in the HPRF cavity is scheduled in 2010. One can expect when high intensity charged beam passes through the HPRF cavity it generates a beam-induced hydrogen plasma in the hydrogen gas. The left hand side of Figure 5 shows the expected beam-loading effect in the RF pickup signal. The RF amplitude is immediately reduced at the beam passing through the cavity since large amount of RF power is absorbed by the electron swarm in the beam-induced plasma. A very slow recombination rate ($10^{-8}$ cm$^3$/s) is used in the simulation. However, no experiment measurement has been taken to determine the recombination rates in such a dense gas condition. One possible expectation is that hydrogen ions ($H^+$ and $H_2^+$) form a polyatomic hydrogen cluster in a very short time (~ps). The reaction chain process is shown in left hand side of Figure 5. The cluster recombines a free electron in a very short time (less than 1 ns). If this process is predominant, the HPRF cavity can survive under high radiation conditions.

Even if the recombination process in pure hydrogen gas is slow, the ionized electrons can be significantly reduced by doping using electronegative gases, i.e. SF6 in the HPRF cavity. In fact, only 0.1 % of SF6 drastically recovers the RF amplitude with high intensity charged beam as shown in left hand side of Figure 5. SF6 is dissociated and forms F$^-$ ions. It is very active and interacts with the electrode metals. Or it can form HF, a toxic compound. Besides, SF6 does not work under cryogenic temperature. Further study is needed.

## SUMMARY

The HCC is being developed to quickly generate exceptional 6D cooling of muon beams. Experimental and simulation approaches have been made to make a practical HCC. The first end-to-end simulation has been done with reasonable lattice parameters and obtained cooling factor $> 10^5$ which is close to the design value for muon colliders. HCC magnet and RF beam elements have been investigated. The helical solenoid magnet has been studied and shown to reproduce the required field components. The calculated field in the helical solenoid is more uniform than the analytical field expressions that have been used up to now in the simulations. The actual HS fields will generate a larger stable beam phase space than the analytical field expressions with better expected performance in beam cooling and RF power requirements. We also demonstrated high pressure hydrogen gas filled RF cavity operation in strong magnet fields. From recent RF breakdown studies, the RF breakdown only happens when the hydrogen plasma density exceeds $10^{18}$ cm$^{-3}$ from spectroscopic measurement of RF breakdown. This value is confirmed from different RF breakdown analysis by using RF pickup signals. The electron flow density is of the order of $10^{14}$ electrons when the breakdown starts. These values are consistent if the size of plasma sheath is 100 μm that is very typical value in the plasma discharge process. The next stage of testing requires operating HPRF cavities in beams of ionizing radiation, which is scheduled to occur in 2010.

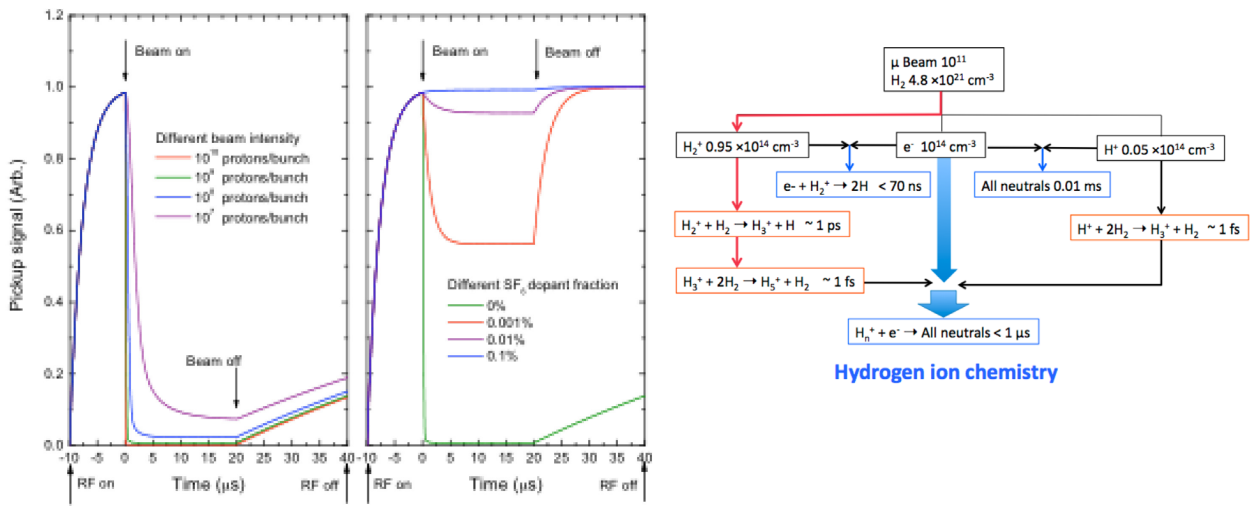

**Figure 5.** (Left) Simulated RF pickup signal and beam loading effects in HPRF cavities. RF cavity Q reduction can be mitigated by doping with an electronegative gas as shown in the right hand side plot where 0.1 % of SF6 almost completely recovers the RF amplitude. (Right) Chain process of hydrogen ions relevant to electron recombination rates.